\documentstyle[prd,aps,epsfig,array,preprint]{revtex}
\newcommand{\al}{\alpha}
\newcommand{\be}{\beta}
\newcommand{\ga}{\gamma}
\newcommand{\la}{\lambda}
\newcommand{\de}{\delta}
\newcommand{\ro}{\rho}
\newcommand{\si}{\sigma}
\newcommand{\na}{\nabla}

\begin{document}
\draft

\title{Stress--energy tensor of neutral massive fields\\
in the Reissner--Nordstr\"om
spacetime}

\author{Jerzy Matyjasek\thanks{Electronic Address:
matyjase@tytan.umcs.lublin.pl, jurek@iris.umcs.lublin.pl}}

\address{Institute of Physics, Maria Curie-Sk\l odowska University,\\
pl. Marii Curie - Sk\l odowskiej 1,\\
20-031 Lublin, Poland}

\maketitle

\begin{abstract}
\noindent
 The approximation of the
 renormalized stress-energy tensor	 of the quantized
 massive scalar, spinor, and vector field   in the Reissner-
 Nordstr\"om spacetime is constructed.
 It is achieved by functional differentiation of the 
 lowest order of the Schwinger-DeWitt effective action involving coincidence limit
of the Hadamard-Minakshisundaram-DeWitt-Seely coefficient $a_{3},$
 and restricting thus obtained general formulas to spacetimes with  vanishing curvature scalar.
 For the massive scalar field with arbitrary curvature coupling	our results
 reproduce those obtained previously by Anderson, Hiscock, and Samuel
 by means of 6-th order WKB approximation.

\end{abstract}

\vskip 0.8cm \noindent {PACS numbers: 04.70.Dy, 04.62+v \\UMCS-TG-99-11}

\preprint{}
% \clearpage
 
 \section{Introduction}
 Treating  the renormalized stress-energy tensor 
 as the source term of the semiclassical Einstein field equations,
 one could, in principle, determine the back reaction of the quantized
 fields upon the spacetime geometry of a black hole
 unless the (unknown) quantum gravity effects become  dominant.	
 Mathematical difficulties encountered in the attempts to
 construct  characteristics of the vacuum polarization in a concrete spacetime are well
 known and since the back reaction equations
 require knowledge of the  functional dependence of stres-energy tensor of the quantized
 field, 
 $  \langle T^{\mu}_{\nu}\rangle_{ren},$ 
 on a wide class of metrics, purely analytical treatment of 
 the problem is impossible.
 It is natural therefore that much effort have been concentrated
 on	developing approximate methods.

 The vacuum polarization effects of the massive fields in the
 curved background has been studied by a number of authors [1-18]. It has been
 shown that for sufficiently massive fields (i.e. when the Compton length
 is much smaller than the characteristic radius of curvature,
 where the latter means any characteristic length scale of the
 geometry) 
 the asymptotic expansion of the effective
 action in powers of $m^{-2}$ may be used. It is because the nonlocal
 contribution to the total effective action can be neglected, and, consequently,
 the vacuum polarization part is local and determined by the geometry
 of the spacetime in question.	 
 
 In the black hole spacetimes the vacuum polarization of the massive 
 scalar, spinor, and vector fields  
 have been constructed	in a series of papers by Frolov and Zel'nikov
 in the vacuum type-D geometries [1-4]. They used general framework of
 the Schwinger - DeWitt method [12-18] and constructed the first order
 of the effective action,
 omitting the terms that do not contribute to a Ricci-flat spaces.
 Using a different method, 
 Anderson, Hiscock, and Samuel evaluated the approximate $ \langle T^{\mu}_{\nu}\rangle_{ren}$ 
 of the massive scalar field with arbitrary curvature coupling
 for a general static, spherically symmetric spacetime and applied 
 obtained formulas to the Reissner-Nordstr\"om spacetime [8].
 Their approximation is equivalent to 
 the Schwinger - DeWitt expansion; to obtain the lowest (i. e. $m^{-2}$) terms,
 one has to use sixth-order WKB approximation of the mode functions. 
 Numerical calculations
 reported in Ref.[8] indicate that the Schwinger-DeWitt method
 always provide a good approximation of the renormalized stress
 energy tensor of the massive scalar field with arbitrary curvature
 coupling as long as the mass of the field remains sufficiently
 large.

 The aim of this paper is to 
 construct the renormalized stress-energy tensor of the massive scalar with
 arbitrary curvature coupling, spinor,
 and vector fields in the geometry of the Reissner-Nordstr\"om black hole.
 To our knowledge the spinor and vector fields have not been discussed earlier.
 We shall achieve this using  the
 standard result of the theory of quantized massive fields in the curved 
 background that connects the coincidence limit of the HDSM 	 
 (Hadamard-Minakshisundaram-DeWitt-Seely)
 coefficient  $[a_{3}]$ with the lowest order of the
 one-loop effective action and consequently
 with the regularized  stress-energy tensor [1-4, 19-23].
 Indeed, functionally differentiating the effective action we obtain  a general and
 rather complex expression for the renormalized stress-energy tensor
 that is valid in any spacetime. Then we specialize thus obtained formulas to the 
 spacetimes with vanishing curvature scalar and apply the result to the Reissner-
 Nordstr\"om geometry.
 We show that for the scalar field the resulting 
 $ \langle T^{\mu}_{\nu}\rangle_{ren}$ is identical
 with the tensor obtained  earlier by Anderson, Hiscock, and Samuel and 
 that in the limit of vanishing electric charge it reduces to the stress-energy
 tensor constructed by Frolov and Zel'nikov.

 There is another important limit of the general Reissner-Nordstr\"om geometry
 that yields the extremal black hole. Expanding the near-horizon region of
 such a geometry  into
 a whole manifold one obtains the Bertotti-Robinson spacetime actively studied
 recently. We construct the stress-energy tensors in the Bertotti-Robinson spacetime 
 taking appropriate limits in our general formulas and analyse the conditions 
 under which this geometry is a self-consistent solution of the semiclassical
 Einstein field equations.  Analyses carried out in the Bertotti-Robinson
 spacetime yield similar results.
 
 The effective action approach that we employ in this paper requires
the metric  of the spacetime to be positively defined.  
Hence, to  obtain the physical stress-energy
tensors one has to analytically continue at the final stage of calculations
their Euclidean counterparts.  
It should be stressed once again that the method, when applied to
the rapidly varying or strong  gravitational fields, breaks down and that its
massless limit is contaminated by nonphysical divergences.

%%%%%%%%%%%%%%%%%%%%%%%%%%%%%%%%% SECTION 2 %%%%%%%%%%%%%%%%%%%%%%%%%%%
\section{The effective action}
We begin with a short description of the method. More detailed
presentation may be found in [3,4, 21-23]. Our notation corresponds to those of
Refs. [21-23].
 Consider the elliptic second-order differential operator of the form
 \begin{equation}
 F = g^{\mu \nu}\na_{\mu}\na_{\nu} \,+\,Q\,-\,m^{2},
 \end{equation}
 acting on the (super)field $\phi^{A}(x),$
 where
 \begin{equation}
 Q\,=\,{Q^{A}}_{B}
 \end{equation}
is some matrix-valued function playing a role of the potential,
$\na_{\mu}$ is the appropriate covariant derivative 
and $m^{2}$ is a matrix satisfying $\nabla_{\mu} m^{2}$	 and commuting with $Q.$
It is unnecessary at this stage to know the exact form of the affine connection;
all that is needed now is the knowledge of
the commutator of the covariant derivatives which defines curvature
according to a rule	
\begin{equation}
[ \na_{\mu},\na_{\nu}]\phi^{A}\,=\, {{\cal R}^{A}}_{B \mu \nu}\phi^{B}.
\end{equation}
 The renormalized effective action constructed from the Green function
 of the differential  operator (1)
 is given by
\begin{equation}
W_{ren}  \,=\,{1\over 32\pi^{2}}
\int g^{1/2} d^{4}x \sum_{n=3}^{\infty}{(n - 3)!\over (m^{2})^{n -2}}{\rm Tr}[a_{n}],
\end{equation}
where 
$[a_{n}]$ is the coincidence limit of the n-th HDSM coefficient and
${\rm Tr}$ is the matrix supertrace	 defined as	 [24]
\begin{equation}
{\rm Tr} = {\rm tr_{+} - tr_{-}},
\end{equation}
where
\begin{equation}
{\rm tr_{\pm}}f = \sum_{A} f_{AA} [ 1 \pm (-1)^{\epsilon_{A}}]{1\over 2}
\end{equation}
and $\epsilon_{A}$ is the Grassman parity of $\phi^{A}.$
The coefficients $a_{0}, a_{1},$ and $a_{2}$ contribute to the divergent
part of the  action , $W_{div},$ which have to be absorbed into the
classical gravitational action
\begin{equation}
S_{g}\,=\,\int g^{1/2} d^{4}x \left( \Lambda_{0}\,+\,\la_{0} R\,+\,
\la_{1} R^{2}\,+\,\la_{2} R^{\mu \nu} R_{\mu \nu}\right)\,+\,\la_{3}\chi,
\end{equation}
where
\begin{equation}
\nonumber
\chi\,=\,\int g^{1/2} d^{4}x \left(R_{\mu \mu\ro \si} R^{\mu \nu \ro \si}\,-\,
4 R_{\mu \nu} R^{\mu \nu}\,+\,R^{2}\right),	
\end{equation}
by renormalization of the bare coupling constant. The  parameters lambda
should be determined experimentally and are expected to be small, since otherwise
they could cause observable deviations from the predictions of the general
relativity. Instead of writing out the squared  Riemann tensor 
we used in (7) 
the  Gauss-Bonnet invariant, which has zero functional derivative
with respect to the metric tensor.

The construction of the coincidence limits of the HDSM coefficients
is, except the first two, an extremely laborious task.
The third coefficient, $[a_{2}],$ that is proportional to the anomalous trace of
the renormalized stress-energy tensor 
of the quantized massless and conformally invariant fields
has been calculated by DeWitt  [12].
The coincidence limit of $a_{3}$ has been obtained by Gilkey [19,20]
whereas the  coefficient $a_{4}$  has been calculated by Avramidi [21-23,25]
and by Amsterdamski et al. [26]. Since we are interested in the lowest order of
the effective action (4) we need simple and general expression for $[a_{3}].$
Here we use $[a_{3}]$ as proposed by Avramidi [21-23] but with a different normalization: 
\begin{equation}
[a_{3}] = {1\over 3!} \left\{ P^{3}\,+\,{1\over 2}\{P, Z_{(2)}\}\,
+\,{1\over 2}\left(\na_{\mu} P\,+\,{1\over 3}J_{\mu}\right)
\left(\na^{\mu} P\,-\,{1\over 3}J^{\mu}\right) \,+\,{1\over 10} Z_{(4)}\right\},
\end{equation}
where
\begin{equation}
J_{\mu}\,=\,\na_{\si}{\cal R^{\si}}_{\mu},
\end{equation}
\begin{equation}
P\,=\,Q\,-\,{1\over 6} \hat{1} R,
\end{equation}
\begin{equation}
Z_{(2)}\,=\,\Box Q\,+\,{1\over 2}{\cal R_{\mu \nu}}{\cal R^{\mu \nu}}\,-\,
\hat{1} \left( {1\over 30} R_{\mu \nu} R^{\mu \nu}\,-\,{1\over 30} R_{\mu \nu \ro \si}
R^{\mu \nu \ro \si}	\,-\,{1\over 5} \Box R \right),
\end{equation}
and
\begin{eqnarray}
Z_{(4)}\,&=&\,\Box^{2} Q\,-\,{1\over 2}[ {\cal R}_{\mu \nu}, [{\cal R^{\mu \nu}}, Q]]\,-\,
{2\over 3}[ J^{\mu}, \na_{\mu} Q]\,+\,{2\over 3} R^{\mu \nu} \na_{\mu}\na_{\nu} Q \,+\,
{1\over 3}\na^{\mu} R \na_{\mu} Q	\nonumber \\ \nonumber
&+&\,2 \{{\cal R_{\mu \nu}}, \na^{\mu} J^{\nu}\}\,+\,{8\over 9} J_{\mu} J^{\mu}	\,+\,
{4\over 3}\na_{\mu}{\cal R_{\ro \si}}\na^{\mu}{\cal R^{\ro \si}}\,+\,
6 {\cal R_{\mu \nu}}{\cal R^{\nu}}_{\ro}{\cal R^{\ro \mu}}\,+\,{10\over 3} R^{\ro \si}
{\cal R^{\mu}}_{\ro}{\cal R_{\mu \si}}\\ \nonumber
&-&\,R^{\mu \nu \ro \si} {\cal R_{\mu \nu}}{\cal R_{\ro \si}}\,-\,
\hat{1}\left( -{3\over 14} \Box^{2} R\,-\,{1\over 7} R^{\mu \nu}\na_{\mu}\na_{\nu} R\,+\,
{2\over 21} R^{\mu \nu} \Box R_{\mu \nu}\,-\,
{4\over 7} {{{R^{\ro}}_{\mu}}^{\si}}_{\nu} \na_{\ro}\na_{\si} R^{\mu \nu}\right.
\\ \nonumber
&-&\,{4\over 63}\na_{\mu} R \na^{\mu} R\,+\,{1\over 42}\na_{\mu} R_{\si \ro}
\na^{\mu} R^{\si \ro}\,+\,{1\over 21}\na_{\mu} R_{\ro \si} \na^{\ro}
R^{\si \mu} \,-\,{3\over 28} \na_{\mu} R_{\ro \si \la \tau} 
\na^{\mu} R^{\ro \si \la \tau}
\\ \nonumber
&-& {2\over 189}{R_{\mu}}^{\nu} {R_{\ro}}^{\mu} {R_{\nu}}^{\ro}\,+\,{2\over 63}
R_{\ro \si} R^{\mu \nu} {{{R^{\ro}}_{\mu} }^{\si}}_{\nu}\,-\,{2\over 9}
R_{\ro \si} {R^{\ro}}_{\mu \nu \la} R^{\si \mu \nu \la}
\,+\,{16\over 189} {R_{\ro \si}}^{\mu \nu} {R_{\mu \nu}}^{\la \ga}
{R_{ \la \ga}}^{\ro \si}
\\ 
 &+& \left.\,{88\over 189} R^{\ro ~ \si}_{~ \mu ~ \nu}
 R^{\mu ~ \nu}_{~ \la ~ \ga} R^{\la ~ \ga}_{~ \ro ~ \si}
 \right).
\end{eqnarray}
In the above formulas $\hat 1$ is the unit matrix, $\left\{\,,\,\right\}$
is the anticommutator
and we have omitted the field indices.

%%%%%%%%%
Inserting (8) into (4)  integrating by parts and making use of the 
elementary properties of the Riemann tensor one has
\begin{eqnarray}
W_{ren}\,&=&\,{1\over 192\pi^{2} m^{2}}\int d^4 x\,g^{1/2}{\rm Tr}\left\{P^3
+{{1}\over {30}}P\left(R_{\mu\nu\alpha\beta}R^{\mu\nu\alpha\beta}
-R_{\mu\nu}R^{\mu\nu}+ \Box R\right)
+{{1}\over {2}}P{\cal R}_{\mu\nu}{\cal R}^{\mu\nu}\right.
\nonumber\\
& &
+{{1}\over {2}} P\Box P -{{1}\over {10}}J_\mu J^\mu
+{{1}\over {30}}
\left(2{\cal R}^\mu_{\ \nu}{\cal R}^\nu_{\ \alpha}{\cal R}^\alpha_{\ \mu}
-2R^\mu_\nu {\cal R}_{\mu\alpha}{\cal R}^{\alpha\nu}
+R^{\mu\nu\alpha\beta}{\cal R}_{\mu\nu}{\cal R}_{\alpha\beta}\right)
\nonumber\\
& &
+\hat 1\left[-{{1}\over {630}}R\Box R
+{{1}\over {140}}R_{\mu\nu}\Box R^{\mu\nu}
+{{1}\over {7560}}\left(
-64R^\mu_\nu R^\nu_\lambda R^\lambda_\mu
+48R^{\mu\nu}R_{\alpha\beta}R^{\alpha\ \beta}_{\ \mu\ \nu}\right.\right.
\nonumber \\
& &
\left. \left. \left.+6R_{\mu\nu}R^\mu_{\ \alpha\beta\gamma}R^{\nu\alpha\beta\gamma}
+17R_{\mu\nu}^{\ \ \ \alpha\beta}
R_{\alpha\beta}^{\ \ \ \sigma\rho}R_{\sigma\rho}^{\ \ \ \mu\nu}
-28R^{\mu \ \nu}_{\ \alpha \ \beta}R^{\alpha\ \beta}_{\ \sigma\ \rho}
R^{\sigma\ \rho}_{\ \mu \ \nu}\right)\right]\right\}.
\end{eqnarray}
This first order renormalized effective action applies to any spacetime and
to any differential operator of the form (1).
In what follows we shall confine ourselves to the operators
\begin{equation}
(-\,\Box \,+\,\xi R \,+\,m^2) \phi^{(0)}\,=0,
\end{equation}
\begin{equation}
(\ga^{\mu} \na_{\mu}\,+\, m) \phi^{(1/2)}\,=\,0,
\end{equation}
\begin{equation}
 (\de^{\mu}_{\nu}\Box\,-\,\na_{\nu}\na^{\mu}\,-
 \,R^{\mu}_{\nu} \,-\,\de^{\mu}_{\nu}m^{2})\phi^{(1)}\,=\,0,
\end{equation}
 acting on the scalar, spinor, and vector fields, respectively,
where $\xi$ is the coupling constant, and $\ga^{\mu}$  are the Dirac matrices
obeying standard relations 
$\ga^{\al} \ga^{\be}\,+\,\ga^{\be} \ga^{\al}\,=\,2 \hat 1 g^{\al \be},$
and assume that the fields are neutral.
Although neither (16) nor (17) has the form that allows direct
application of the Schwinger-DeWitt technique the procedures described in
Refs [3,24]  may be used in this context. Specifically,
by appropriate redefinition of the massive spinor field
one obtains
\begin{equation}
\left(\nabla_{\mu} \nabla^{\nu}\,-\,{1\over 4} R\,-\,m^{2}\right) \phi^{(1/2)}\,=\,0,
\end{equation}
whereas the method presented in Refs.[24] shows that the effective
action of the massive vector field is equal to the effective action of the 
diagonal operator
\begin{equation}
(\de^{\mu}_{\nu}\Box\,-
 \,R^{\mu}_{\nu} \,-\,\de^{\mu}_{\nu}m^{2})\phi^{(1)}\,=\,0,
\end{equation}
minus the effective action of the massive scalar field with the minimal curvature coupling.
The first order of the effective action is therefore
\begin{equation}
W^{(1)}_{ren}\,=\, {1\over 32 \pi^{2} m^{2}}\int g^{1/2} d^{4}x\left\{\begin{array}{l}%\bigskip
[a^{(0)}_{3}]\\
%\bigskip
-tr [a^{(1/2)}_{3}]\\
%\bigskip
tr [a^{(1)}_{3}]\,-\,[a^{(0)}_{3|\xi = 0}]
\end{array}\right.
\end{equation}
where the definition of the matrix supertrace has been used.

For fields obeying Eqs.(15-17) the curvature  has the form
\begin{equation}
 {\cal R_{\mu \nu}}\,=\,\left\{\begin{array}{l}%\bigskip
0\\
%\bigskip
{1\over 4}\ga^{\ro}\ga^{\si} R_{\ro \si \mu \nu}\\
%\bigskip
{R^{\ro}}_{\si \mu \nu}
\end{array}\right. 
\end{equation}
whereas inspection of (15), (17) and (18) shows that the potential matrix is

\begin{equation}
 Q\,=\,\left\{\begin{array}{l}%\bigskip
- \xi R\\
%\bigskip
- {1\over 4}\hat{1} R\\
%\bigskip
-{R^{\al}}_{\be}
\end{array}\right.
\end{equation}
Inserting (21) and (22) into (14)  making use of elementary properties of
the Dirac matrices	and Riemann tensor, after 
simple calculations one obtains the  
first term of the asymptotic expansion of the effective action in the form [21,23]
 \begin{eqnarray}
 W^{(1)}_{ren}\,&=&\,{1\over 192 \pi^{2} m^{2}} \int d^{4}x g^{1/2}
 \left( \al^{(s)}_{1} R \Box R\,+\,\al^{(s)}_{2} R_{\mu \nu} \Box R^{\mu \nu}\,+\,
 \al^{(s)}_{3} R^{3}\,+\,\al^{(s)}_{4} R R_{\mu \nu} R^{\mu \nu}	  \right. \\ \nonumber
 &+&\,\al^{(s)}_{5} R R_{\mu \nu \ro \si} R^{\mu \nu \ro \si}\,+\,
 \al^{(s)}_{6} R^{\mu}_{\nu} R^{\nu}_{\ro} R^{\ro}_{\mu}\,+\,\al^{(s)}_{7} R^{\mu \nu}
 R_{\ro \si} R^{\ro ~ \si}_{~ \mu ~ \nu} \\ \nonumber
&+&\left. 
\,\al^{(s)}_{8} R_{\mu \nu} R^{\mu}_{\la \ro \si} R^{\nu \la \ro \si}\,+\,
\al^{(s)}_{9} {R_{\ro \si}}^{\mu \nu} {R_{\mu \nu}}^{\la \ga}
{R_{ \la \ga}}^{\ro \si}
\,+\,\al^{(s)}_{10}
R^{\ro ~ \si}_{~ \mu ~ \nu}
 R^{\mu ~ \nu}_{~ \la ~ \ga} R^{\la ~ \ga}_{~ \ro ~ \si},
 \right) \\ \nonumber
 &=&\,{1 \over 192 \pi^{2} m^{2}} \sum_{i=1}^{10} \al^{(s)}_i W_{i},
\end{eqnarray}
where the numerical coefficients depending on the spin of the field are  given in a Table I.
Note that because of (20) our coefficients $\al^{(1/2)}_{i}$ for the spinor field are twice
these of Avramidi's~\cite{avra1}, and
to obtain the result for the massive neutral spinor field on has to multiply
$W^{(1)}_{ren}$ by the factor $1/2.$

\section{The stress-energy tensor in $R\,=\,0$ geometries}
The renormalized stress-energy tensor is given by
\begin{equation}
{2\over g^{1/2}}{\delta\over \delta g_{\mu \nu}} W^{(1)}_{ren}\,=\, 
\langle T^{\mu \nu}\rangle_{ren},
\end{equation}
and for the massive scalar, spinor, and vactor fields
may be rewritten in terms of the variational derivatives of the actions $W_{i}$ in the form
\begin{equation}
  \langle T^{\mu\nu}\rangle_{ren}^{(s)}\,=\,{1\over 96 \pi^{2} m^{2}}{1\over g^{1/2} }
\sum_{i=1}^{10}\al^{(s)}_{i} {\delta W_{i}\over \delta g_{\mu \nu}}.
\end{equation}
Functionally differentiating the renormalized effective action with respect to the
 metric tensor, performing elementary simplifications
 and finally retaining in the  result only the terms that are nonzero
 for $R\,=\,0$ geometries, after rather long calculations, one has

 %%%%%%%%%%%%%%%%%%%%%%%%%%%%%%%%%%%%%%%%%
\begin{equation}
{1\over g^{1/2}}{\de\over \de g_{\mu \nu}}	W_{1}=(...),
\end{equation}
\begin{eqnarray}
{1\over g^{1/2}}{\de\over \de g_{\mu \nu}}W_{2}&=&
\na^{\mu} R_{\ro \si} \na^{\nu} R^{\ro \si}\,+\,
\na^{\mu} R_{\ro \si} \na^{\si} R^{\ro \nu}\,-\,
3 \na^{\mu} R_{\ro \si} \na^{\si} R^{\ro \nu} \nonumber \\
&+&\,
2 \na^{\ro}\na^{\nu}\Box R_{\ro}^{~\mu}
\,-\,
\Box^{2} R^{\mu \nu}\,-\,{1\over 2} \na_{\la} R_{\ro \si} \na^{\la} R^{\ro \si}	g^{\mu \nu}
\,-\,\na^{\ro} \na^{\si} \Box R_{\ro \si} g^{\mu \nu}\nonumber \\ 
&+&\,3 \na_{\si} \na^{\nu} R_{\ro}^{~ \mu} R^{\ro \si}\,-\,
\na_{\si} \na^{\mu} R_{\ro}^{~ \nu} R^{\ro \si}
\,-\,\Box R_{\ro}^{~ \nu} R^{\ro \mu}
\,-\,3 \na^{\si} \na^{\mu} R_{\ro \si} R^{\ro \nu} \nonumber \\
&-&\,\Box R_{\ro}^{~ \mu} R^{\ro \nu}
\,+\,\na^{\ro} \na^{\nu} R_{\ro \si} R^{\si \mu} \,
+\, (...),
\end{eqnarray}
\begin{equation}
{1\over g^{1/2}}{\de\over \de g_{\mu \nu}}W_{3}= (...),
\end{equation}

\begin{eqnarray}
{1\over g^{1/2}}{\de\over \de g_{\mu \nu}} W_{4}&=&
2 \na^{\mu} R_{\ro \si} \na^{\nu} R^{\ro \si}\,-\,2 \na_{\la} R_{\ro \si} \na^{\la}R^{\ro \si}
g^{\mu \nu}\,+\,2 \na^{\nu}\na^{\mu} R_{\ro \si} R^{\ro \si}\nonumber \\
&-&\,2 \Box R_{\ro \si} R^{\ro \si} g^{\mu \nu}\,-\,R_{\ro \si} R^{\ro \si} R^{\mu \nu}	\,+\,(...),
\end{eqnarray}

\begin{eqnarray}
{1\over g^{1/2}}{\de\over \de g_{\mu \nu}}
W_{5}&=&
%&&\mn\int R R_{\al \be \la \ga}\Box R^{\al \be \la \ga} g^{1/2} d^{4}x\,
2 \na^{\mu} R_{\ro \si \la \ga}\na^{\nu} R^{\ro \si \la \ga}\,-\,
2 \na_{\tau} R_{\ro \si \la \ga} \na^{\tau} R^{\ro \si \la \ga} g^{\mu \nu}	\nonumber \\
&+&\,
2 \na^{\mu}\na^{\nu} R_{\ro \si \la \ga} R^{\ro \si \la \ga}
\,-\,2\Box R_{\ro \si \la \ga} R^{\ro \si \la \ga} g^{\mu \nu}\,
-\,R^{\mu \nu}
R_{\ro \si \la \ga} R^{\ro \si \la \ga}\,+\,(...),
\end{eqnarray}

\begin{eqnarray}
{1\over g^{1/2}}{\de\over \de g_{\mu \nu}}	W_{6}&=&
%&&\mn\int R^{\al}_{~ \be} R^{\be}_{~ \ro} R^{\ro}_{~ \al} g^{1/2} d^{4}x
3 \na^{\nu} R_{\ro \si}\na^{\si} R^{\ro \mu}\,-\,3 \na_{\si} R_{\ro}^{~ \nu}\na^{\nu} R^{\si \mu}
\,-\,
{3\over 2} \na_{\la} R_{\ro \si} \na^{\si} R^{\ro \la} g^{\mu \nu}
\nonumber\\
&+&\,
3 \na_{\si} \na_{\nu} R_{\ro}^{~ \mu} R^{\ro \si}
-{3\over 2} \na^{\si} \na_{\la} R_{\ro \si} R^{\ro \la} g^{\mu \nu}\,+\,
3 \na^{\si} \na^{\nu} R_{\ro \si} R^{\ro \mu}
\,-\,{3\over 2}\Box R_{\ro}^{~\nu}	R^{\ro \mu}	\nonumber \\
&-&\,3 R_{\ro}^{~ \si} R_{\si}^{~ \nu} R^{\ro \mu}
-\,{3\over 2} \Box R_{\ro}^{~ \mu} 
R^{\ro \nu}
\,+\,{1\over 2} g^{\mu \nu} R_{\ro \si} R_{\la}^{~ \ro} R^{\si \la},
\end{eqnarray}

\begin{eqnarray}
{1\over g^{1/2}}{\de\over \de g_{\mu \nu}} W_{7}&=&
%&&\mn\int R^{\al \be} R_{\ro \si} R^{\ro ~ \si}_{~ \al ~ \be}
%g^{1/2} d^{4}x\,=\,
\na_{\si} R_{\ro}^{~ \mu} \na^{\ro} R^{\si \nu}\,-\,
2 \na^{\nu} R_{\ro \si}\na^{\la} R_{\la}^{~ \ro \si \mu}\,+\,
2 \na_{\la} R_{\ro \si} \na^{\nu} R^{\ro \la \si \mu} \nonumber \\
&-&\,2 \na_{\la} R_{\ro \si} \na^{\la} R^{\ro \mu \si \nu} 
\,+\,
2 \na_{\la} R_{\ro \si} \na^{\ga} R_{\ga}^{~ \ro \si \la}\,-\,
\na_{\si}\na_{\ro} R^{\mu \nu} R^{\ro \si}
\,+\,2 \na^{\la} \na^{\nu} R_{\ro ~ \si \la}^{~ \mu} R^{\ro \si}\nonumber \\
&-&\,
\Box R_{\ro ~ \si}^{~ \mu ~ \nu} R^{\ro \si}
\,-\,
\na^{\si}\na^{\ga} R_{\ro \si \la \ga} R^{\ro \la} g^{\mu \nu}
+\,{1\over 2} \na^{\ro} \na_{\si} R_{\ro}^{~ \nu} R^{\si \mu}\,+\,
{1\over 2} \na^{\ro} \na_{\si} R_{\ro}^{~ \mu} R^{\si \nu}\nonumber \\
&+&\,
{1\over 2} R_{\ro \si} R_{\la \ga} R^{\ro \la \si \ga}
\,+\,2 \na_{\la} \na^{\nu} R_{\ro \si} R^{\ro \la \si \mu}\,-\,
3 R_{\ro \si} R_{\la}^{~ \mu} R^{\ro \la \si \nu}\,-\,
\na_{\ga} \na_{\la}	R_{\ro \si} R^{\ro \ga \si \la} g^{\mu \nu}\nonumber \\
&-&\,\Box  R_{\ro \si} R^{\ro \mu \si \nu}, 
\end{eqnarray}

\begin{eqnarray}
{1\over g^{1/2}} {\de\over \de g_{\mu \nu}}W_{8}&=&
%&&\mn\int R_{\al \be} R^{\al}_{\ro \si \la}
%R^{\be \ro \si \la}
%g^{1/2} d^{4}x  \,=\,
-2 \na_{\si} R_{\ro}^{~ \mu} \na^{\la} R_{\la}^{~ \ro \si \nu}\,+\,
\na_{\si} R_{\ro}^{~ \mu} \na^{\la} R_{\la}^{~ \nu \ro \si}\nonumber \\
&+&\,
\na^{\nu} R_{\ro \si \la \ga}\na^{\ga} R^{\ro \si \la \mu}
\,-\,\na_{\ga} R_{\ro \si \la}^{~ ~ ~ \mu} \na^{\ga}R^{\ro \si \la \mu}
\,+\,
\na^{\si} R_{\ro \si \la \ga} \na^{\nu} R^{\ro \mu \la \ga}	\nonumber \\
&-&\,
2 \na_{\la} R_{\ro \si} \na^{\si} R^{\ro \mu \la \nu}
\,-\,{1\over 2}\na^{\ro} R_{\ro \si \la \ga} \na^{\tau}R_{\tau}^{~ \si \la \si} g^{\mu \nu}
+\,{1\over 2} \na_{\tau}R_{\ro \si \la \ga} \na^{\la} R^{\ro \si \ga \tau} g^{\mu \nu}
\nonumber \\
&-&\,
2 \na_{\la} \na^{\ro} R_{\ro ~ \si}^{~ \nu ~ \mu} R^{\si \la}
-\,  2 \na^{\ro} \na^{\la} R_{\ro \si \la}^{~ ~ ~ \nu} R^{\si \mu}
\,+\,
\na^{\nu}\na_{\ga} R_{\ro ~ \si \la}^{~ \mu} R^{\ro \ga \si \la}
\nonumber \\
&+&\,
R_{\ro}^{~ \mu} R_{\si \la \ga}^{~ ~ ~ \nu} R^{\ro \ga \si \la}
\,-\,{1\over 2}\na_{\tau}\na^{\si} R_{\ro \si \la \ga} R^{\ro \tau \la \ga} g^{\mu \nu}\,-\,
{1\over 2}\Box R_{\ro ~ \si \la}^{~ \nu} R^{\ro \mu \si \la}\nonumber \\
&+&\,
\na^{\si}\na^{\nu} R_{\ro \si \la \ga} R^{\ro \mu \la \ga}
\,-\,\,{1\over 2} \Box R_{\ro ~ \si \la}^{~ \mu} R^{\ro \nu \si \la}
\,+\,
2 R_{\ro \si} R_{\la ~ \ga}^{~ \ro ~ \mu} R^{\si \la \ga \nu}
\nonumber \\
&+&\,
{1\over 2} \na_{\tau}\na^{\ro} R_{\ro \si \la \ga} R^{\si \tau \la \ga} g^{\mu \nu}
\,-\,{1\over 2} R_{\ro \si} R_{\la \ga \tau}^{~ ~ ~ \ro} R^{\si \tau \la \ga}\,-\,
2 \na^{\ro}\na_{\la} R_{\ro \si} R^{\si \mu \la \nu}
\nonumber \\
&+&\,
\na_{\la}\na_{\si} R_{\ro}^{~ \mu} R^{\ro \la \si \nu},
\end{eqnarray}

\begin{eqnarray}
 {1\over g^{1/2}}{\de\over 
 \de g_{\mu \nu}} W_{9}&=&
 %&&\mn\int R_{\al \be}^{~ ~ \ro \si} R_{\ro \si}^{~ ~ \la \ga}
 %R_{\la \ga}^{~ ~ \al \be}
 %g^{1/2} d^{4}x
 - 6 \na^{\la} R_{\ro \si \la}^{~ ~ ~ \nu} \na^{\ga}
R_{\ga}^{~ \mu \ro \si}\,-\,6 \na_{\ga} R_{\ro \si \la}^{~ ~ ~ \nu}
\na^{\la}R^{\ro \si \ga \mu}
\nonumber \\
&-&\,
3 R_{\ro \si \la}^{~ ~ ~ \mu} R_{\ga \tau}^{~ ~ \la \nu} R^{\ro \si \ga \tau}\,
-\,2 \na^{\la} \na_{\ga} R_{\ro \si \la}^{~ ~ ~ \nu} R^{\ro \si \ga \mu}
\,-\,
4 \na^{\la}\na_{\ga} R_{\ro \si \la}^{~ ~ ~ \mu} R^{\ro \si \ga \nu}
\nonumber \\
&-&\,
4 \na_{\ga}\na^{\ro} R_{\ro ~ \si \la}^{~ \mu} R^{\si \la \ga \mu}
\,-\,2 \na_{\ga}\na^{\ro} R_{\ro ~ \si \la}^{~ \mu} R^{\si \la \ga \nu}
\,+\,
{1\over 2} R_{\ro \si \la \ga} R_{\tau \kappa}^{~ ~ \ro \si} R^{\la \ga \tau \kappa}  g^{\mu \nu},
\nonumber \\
\end{eqnarray}
 and
\begin{eqnarray}
{1\over g^{1/2}} {\de\over \de g_{\mu \nu}} W_{10}&=&
%&&\mn\int R^{\al ~ \be}_{~ \ro ~ \si} R^{\ro ~ \si}_{~ \la ~ \ga}
% R^{\la ~ \ga}_{~ \al ~ \be}
%g^{1/2} d^{4}x  \,=\,
3 \na^{\ro} R_{\ro \si \la}^{~ ~ ~ \mu}\na^{\ga}R_{\ga}^{~ \la \si \nu}\,+\,
3 \na_{\ga}	R_{\ro \si \la}^{~ ~ ~ \nu}\na^{\si} R^{\ro \mu \la \ga}
\nonumber \\
&+&\,
3 \na^{\ro} R_{\ro \si \la \ga} \na^{\ga} R^{\si \mu \la \nu}\,
+\,3 \na^{\ro} R_{\ro \si \la \ga} \na^{\ga} R^{\si \nu \la \mu}
\,-\,
3 \na_{\ga} \na_{\la} R_{\ro ~ \si}^{~ \mu ~ \nu} R^{\ro \ga \si \la}
\nonumber \\
&+&\,
\na_{\ga}\na^{\la} R_{\ro ~ \si \la}^{~ \nu} R^{\ro \ga \si \mu} 
\,+\, 2 \na_{\ga} \na^{\la} R_{\ro ~ \si \la}^{~ \mu} R^{\ro \ga \si \nu}\,+\,
3 R_{\ro \si \la \ga} R_{\tau}^{~ \si \ga \mu} R^{\ro \tau \la \nu}
\nonumber \\
&+&\,
2 \na^{\si} \na_{\ga} R_{\ro \si \la}^{~ ~ ~ \nu} R^{\ro \mu \la \ga}
\,-\,3 \na^{\si} \na^{\ga} R_{\ro \si \la \ga} R^{\ro \mu \la \nu}
\,+\,
\na^{\si} \na_{\ga}	R_{\ro \si \la}^{~ ~ ~ \mu} R^{\ro \nu \la \ga}
\nonumber \\
&-&\,
{1\over 2} R_{\ro \si \la \ga} R_{\tau ~ \kappa}^{~ \ro ~ \ga} R^{\si \tau \la \kappa} 
g^{\mu \nu},
\end{eqnarray}
%%%%%%%%%%%%%%%%%%%%%%%%%%%%%%%%%%%%%%%%%%%%%%%%%%%%%%%%%
where the ellipsis denote  omitted terms that contain the scalar curvature
and its covariant derivatives. For  nondiagonal $R = 0$ metrics obtained
 $\langle T^{\mu}_{\nu}\rangle_{ren}^{(s)}$  must be symmetrized.
As expected, the functional derivatives $W_{1}$ and $W_{3}$
do not contribute to
the stress-energy tensor in the Reissner-Nordstr\"om spacetime. 
It should be noted that Eqs ( 26-35  )
have been obtained	by putting $R\,=\,0$  in the general result,
which is more complex and shall not be presented here.
Inspection of Eqs. (26-35) shows that to construct the stress-energy tensor 
of the massive fields in the Ricci-flat
geometry it suffices to analyse only $W_{5},$ $W_{8},$ $W_{9},$ $W_{10},$
and therefore our $\langle T^{\mu}_{\nu}\rangle_{ren}^{(s)}$  generalizes
earlier results derived by Frolov and Zel'nikov [1-3].

%%%%%%%%%%%%%%%%%%%%%%%%% Section IV %%%%%%%%%%%%%%%%%%%%%%%%%%%%%%%%%%%%%%%

\section{ $\langle T^{\mu}_{\nu}\rangle^{(s)}_{ren}$ in the Reissner-Nordstr\"om
spacetime}
Now we are ready to construct the stress-energy tensor of
the massive quantized fields in the nonextremal Reissner-Nordstr\"om geometry
described by the line element
\begin{equation}
d s^{2}\,=\,-\left( 1 - {2 M\over r} + {e^{2}\over r^{2}} \right) d t^{2}\,+\,
 \left( 1 - {2 M\over r} + {e^{2}\over r^{2}} \right)^{-1} d r^{2}\,+\,
 r^{2}\left( \sin^{2}\theta d\phi^{2}\,+\,d\theta^{2}\right),
 \end{equation}
 where $M$ is the mass and $e$ 
 is a charge of the black hole.
 For $e^{2} < M^{2}$ the equation $g_{00}\,=\,0$ has two positive roots
 \begin{equation}
   r_{\pm}\,=\,M\,\pm\,(M^{2}\,-\,e^{2})^{1/2},
\end{equation}
 and the larger root represents the location of the event horizon whereas $r_{-}$
 is the inner horizon.
 In the limit $ e^{2} = M^{2}$ horizons merge at $r = M,$ and the the Riessner-Nordstr\"om
 solution degenerates to the extremal one with 
the line element given by
\begin{equation}
d s^{2}\,=\,\left( 1 - { M\over r}  \right)^{2} d t^{2}\,+\,
 \left( 1 - { M\over r}  \right)^{-2} d r^{2}\,+\,
 r^{2}\left( \sin^{2}\theta d\phi^{2}\,+\,d\theta^{2}\right).
 \end{equation}
 
 Although the stress-energy tensor could be evaluated in the nonstatic
 background (provided the changes of the  geometry are slow)
 we shall confine ourselves to the exterior region
where the spacetime is static. 
%%%%%%%%%%%%%%%%%%%%%%%%%%% Scalar field %%%%%%%%%%%%%%%%%%%%%%%%%%%%%
We begin with the massive scalar field extensively studied in the Ref.[8].
Constructing components of the  Riemann tensor, its contractions and covariant derivatives 
and subsequently  inserting  them with appropriate 
coefficients $\al_{i}^{(0)}$ into (25) we have

\begin{equation}
  \langle T^{\mu}_{\nu}\rangle_{ren}^{(0)}\,=\,C^{\mu}_{\nu}\,+\,\left(\xi	\,-\,{1\over 6}\right) D^{\mu}_{\nu},
\end{equation}
where
 \begin{eqnarray}
 C^{t}_{t}\,&=&\,-{\frac {1}{30240\,\pi^{2}\,m^{2}\, r^{12}}}\,\left( 1248\,{e}^{6}-810\,
 {r}^{4}{e}^{2}+855\,{M}^{2}{r}^{4}+202\,{r}^{2}{e}^{4}  \right. \nonumber \\
&-&\left. \,1878\,{M}^{3}{r}^{3}+1152\,M{r}^{3}{e
}^{2}+2307\,{M}^{2}{r}^{2}{e}^{2}-3084\,r M{e}^{4}\right),
 \end{eqnarray}

 % Maple output
\begin{eqnarray}
D^{t}_{t}\,&=&\,\frac {1}{720\,\pi^{2}\,m^{2}\,r^{12}}\,\left (-792\,{M}^{3}{r}^{3}
+360\,{M}^{2}{r}^
{4}+2604\,{M}^{2}{r}^{2}{e}^{2}\right.\nonumber \\
&-&\left.\,1008\,M{r}^{3}{e}^{2} \,-
\,2712\,r M{e}^{4}+
819\,{e}^{6}+728\,{r}^{2}{e}^{4}\right ),
\end{eqnarray}

% Maple output
\begin{eqnarray}
C^{r}_{r}\,&=&\,\frac {1}{30240\, \pi^{2}\,m^{2}\,r^{12}}\left( 444\,{e}^{6}
-1488\,M{r}^{3}{e}^{2}+162\,{r
}^{4}{e}^{2}
+\,842\,{r}^{2}{e}^{4}-1932\,r M{e}^{4}\right. \nonumber \\
&+&\left.\,315\,{M}^{2}{r}^{4}+
2127\,{M}^{2}{r}^{2}{e}^{2}-462\,{M}^{3}{r}^{3}\right),
\end{eqnarray}

\begin{eqnarray}
 D^{r}_{r}\,&=&\,
 \frac {1}{720\,\pi^{2}\,m^{2}\,r^{12}}\,\left (-792\,{M}^{3}{r}^{3}+360\,{M}^{2}{r}^
{4}+2604\,{M}^{2}{r}^{2}{e}^{2}-1008\,M {r}^{3}{e}^{2}\right. \nonumber \\
&-&\left.\,
2712\,r M{e}^{4}+
819\,{e}^{6}+728\,{r}^{2}{e}^{4}\right ),
\end{eqnarray}

\begin{eqnarray}
 C^{\theta}_{\theta}\,&=&\,-\,\frac {1}{30240\,\pi^{2}\,m^{2}\,r^{12}}
 \,\left( 3044\,{r}^{2}{e}^{4}-2202\,{M}^{3}{r}^{3}
-10356\,r M {e}^{4} \right. \nonumber \\
&+&\left.\,
3066\,{e}^{6}-4884\,{r}^{3}M{e}^{2}+9909\,{r}^{2}{M}
^{2}{e}^{2}+945\,{M}^{2}{r}^{4}+486\,{r}^{4}{e}^{2}\right),
\end{eqnarray}
and
\begin{eqnarray}
D^{\theta}_{\theta}\,&=&\,
\frac {1}{720\,\pi^{2}\,m^{2}\,r^{12}}\,\left (3276\,{r}^{2}{M}^{2}{e}^{2}-1176\,{r}
^{3}M{e}^{2}-3408\,r M{e}^{4}+1053\,{e}^{6}\right.\nonumber \\
&-&\left. \,1008\,{M}^{3}{r}^{3}+432\,{
M}^{2}{r}^{4}+832\,{r}^{2}{e}^{4}\right ).
\end{eqnarray}
Obtained result for nonvanishing components of the stress-energy tensor
coincides with the $  \langle T^{\mu}_{\nu}\rangle_{ren}$ constructed by Anderson, Hiscock
and Samuel.	 This coincidence is, of course, not surprising as there is
a one-to-one correspondence between the order of the WKB approximation
and the order of the Schwinger-DeWitt expansion. 
To obtain the $m^{-2}-$terms one has to use 6-th order WKB approximation of the mode
functions and 	the results (39-45) are simply manifestation
of this correspondence.

%%%%%%%%%%%%%%%% Spinor field %%%%%%%%%%%%%%%%%%%%%%%%%%%%%%%%%%%%%%%%%%%%%%%%
Having computed functional derivatives of $W_{i}$ with respect 
to the metric tensor the construction of the stress-energy tensor 
of the massive fields of higher spins present no problems.
Indeed, inserting  coefficients $\al^{(1/2)}_{i}$ for the neutral spinor field
into (25) one obtains
\begin{eqnarray}
 \langle T^{t}_{t}\rangle^{(1/2)}_{ren}\,&=&\,
\frac {1}{40320\,\pi^{2}\,m^{2}\,r^{12}}\,\left( 2384\,{M}^{3}{r}^{3}+10544\,{r}^{2}{e}^{4}
-22464\,{r}^{3}M {e}^{2}+21832\,{r}^{2}{M}^{2}{e}^{2} \right. \nonumber \\
&-&\left. \,1080\,{M}^{2}{r}^
{4}-21496\,r M{e}^{4}+4917\,{e}^{6}+5400\,{r}^{4}{e}^{2}\right),
\end{eqnarray}

\begin{eqnarray}
  \langle T^{r}_{r}\rangle^{(1/2)}_{ren}\,&=&\,  \frac {1}{40320 \pi^{2} \,m^{2}\, r^{12}}
\,\left( 504\,{M}^{2}{r}^{4}+1080\,{r}^{4}{e}^{2}-
784\,{M}^{3}{r}^{3}-6336\,{r}^{3}M{e}^{2}\right.\nonumber \\
&+&\left. \,3560\,{r}^{2}{e}^{4}+8440\,{
r}^{2}{M}^{2}{e}^{2}-8680\,r M {e}^{4}+2253\,{e}^{6}\right),
\end{eqnarray}
and
\begin{eqnarray}
  \langle T^{\theta}_{\theta}\rangle^{(1/2)}_{ren}\,&=&\, -\frac {1}{40320\,\pi^{2}\,m^{2} \,r^{12}}
\left(-3536\,{M}^{3}{r}^{3}+12080\,{r}^{2}{e}^{
4}-20016\,{r}^{3}M{e}^{2}+30808\,{r}^{2}{M}^{2}{e}^{2}\right.\nonumber \\
&+&\left.\,1512\,{M}^{2}{r
}^{4}-33984\,rM{e}^{4}+9933\,{e}^{6}+3240\,{r}^{4}{e}^{2}\right).
\end{eqnarray}

%%%%%%%%%%%%%%%%%%%%%%%% Vector field %%%%%%%%%%%%%%%%%%%%%%%%%%%%%%%%%%%%%%%%
Similarly, repeating the steps  for the massive vector field one gets
\begin{eqnarray}
  \langle T^{t}_{t}\rangle^{(1)}_{ren}\,&=&\,-\frac {1}{10080 \,\pi^{2}\,m^{2}\,r^{12}}\,\left(
-31057\,{e}^{6}-1665\,{M}^{2}{r}^{4}-
41854\,{r}^{2}{e}^{4}-93537\,{r}^{2}{e}^{2}{M}^{2}\right.\nonumber \\
&+&\left.\,107516\,r M {e}^{4}+
3666\,{M}^{3}{r}^{3}+69024\,{r}^{3}{e}^{2}M-12150\,{e}^{2}{r}^{4}\right),
\end{eqnarray}

\begin{eqnarray}
  \langle T^{r}_{r}\rangle^{(1)}_{ren}\,&=&\,
\frac {1}{10080 \,\pi^{2}\,m^{2}\,r^{12}}\,\left( 1050\,{M}^{3}{r}^{3}-693\,{M}^{2}{r}^{4}+
12907\,{r}^{2}{e}^{2}{M}^{2}-10448\,{r}^{3}{e}^{2}M	\right.\nonumber \\
&-&\left.\,
16996\,r M {e}^{4}+
2430\,{e}^{2}{r}^{4}+6442\,{r}^{2}{e}^{4}+5365\,{e}^{6}\right),
\end{eqnarray}
and
\begin{eqnarray}
  \langle T^{\theta}_{\theta}\rangle^{(1)}_{ren}\,&=&\,
-\,\frac {1}{10080\,\pi^{2}\,m^{2}\,r^{12}}\,\left( 13979\,{e}^{6}-2079\,{M}^{2}{r}^{4}+20908
\,{r}^{2}{e}^{4}+30881\,{r}^{2}{e}^{2}{M}^{2}\right.\nonumber \\
&-&\left.\,
44068\,r M {e}^{4}+4854\,{
m}^{3}{r}^{3}-31708\,{r}^{3}{e}^{2} M +7290\,{e}^{2}{r}^{4}\right).
\end{eqnarray}
Simple calculations show that the tensors (46-48) and (49-51) are covariantly conserved.
Moreover,  it could be easily verified that taking the limit $e = 0$
results, as expected, in the formulas derived by Frolov and Zel'nikov in
the Schwarzschild spacetime (see for 
example Ref. [5]  ).
Although there are no numeric calculations of the stress-energy tensor
of the massive spinor and vector fields against which 
one could test the  results (46-51),
we expect that the approximation is reasonable  so long the mass
of the field is sufficiently large.

Since the Schwinger-DeWitt approximation is local and the 
geometry at the event horizon is regular, one expects that
the stress-energy tensor is	also regular there.
Indeed, it could be easily shown that if there are no fluxes
of energy the regularity conditions on the event horizon [27,28]
require that the components of the 
stress-energy tensor 
and 
\begin{equation}
 \left( 1 - {2 M\over r} + {e^2\over r^2 }\right)^{-1/2}\left(
 \langle T^{t}_{t}\rangle^{(0)}_{ren}\,-\, \langle T^{r}_{r}\rangle^{(0)}_{ren}\right)
\end{equation}
remain finite as $r \to r_{+}.$
Since the difference between the time and radial components of the stress-energy tensor
factors, i.e.
\begin{equation}																				 
  \langle T_{t}^{t}\rangle^{(s)}_{ren}\,-\, 
  \langle T_{r}^{r}\rangle^{(s)}_{ren}\,=\,\left(1 - {2 M\over r} +
{e^{2}\over r^{2}}\right) 
F^{(s)}(r),
\end{equation}
where $F^{(s)}(r)$ 	for each spin of the field is a simple polynomial in $1/r$,
one can draw a general conclusion that our approximate
the stress-energy tensors are regular 
as one approaches the event horizon.	
Analyses carried out in Ref.[8] indicate that all components of
the numerically evaluated stress-energy tensor of the massive scalar field are
also finite on the event horizon.
Repeating calculations in the spacetime of the extreme Reissner-Nordstr\"om black hole,
one obtains the stress-energy tensor that is  regular on the event horizon.
Simple calculations show that
\begin{equation}
\left( 	   \langle T^{t}_{t}\rangle^{(s)}_{ren}\,-\,
  \langle T^{r}_{r}\rangle^{(s)}_{ren} \right)\left( 1 - {M\over r}\right)^{-2}
\end{equation}
is finite  at $r = M.$

To study  obtained  
$  \langle T_{\mu}^{\nu}\rangle^{(s)}_{ren}$ further it is useful to introduce
new coordinate $x\,=\,(r - r_{+})/M$ and a new  parameter
$q = |e|/M.$ 
Since for the massive scalar field Anderson, Hiscock, and Samuel have 
found that for $ m \geq 2/M $ the 
Schwinger-DeWitt approximation is rather good near the event horizon we also take this
bound in our calculations of the vacuum polarization of the spinor and vector fields. 
Our results for $q =0,$ $q = 0.95,$ and $ m \,M = 2$
are displayed in the Figures  1-6. 
Inspection of the  figures shows that for $q$ close to the extremal value, 
the radial dependence of the components of the stress-energy tensor of the 
massive spinor field and their vector counterparts is qualitatively similar.
On the other hand, for small $q$  only the radial components exhibits 
such a similarity. Indeed, on the event horizon $  \langle T^{t}_{t}\rangle^{(1)}_{ren}$ and $  \langle T^{t}_{t}\rangle^{(1/2)}_{ren}$
differ in sign. The difference in the sign of the horizon
values of the stress-energy tensor occurs also  for the tangential components.
Moreover, in the vicinity of the event horizon the magnitude  of the vacuum polariation
effects increases with spin of the quantized field.

Geometries that could be obtained from
nonextremal black holes	taking the extremality  limit
and expanding the  near-horizon geometry into the whole manifold
received recently much attention. 	
Near the event horizon of the extremal Reissner-Nordstr\"om black hole
the geometry approaches that of the  Bertotti-Robinson [29]
\begin{equation}
ds^{2}\,=\,{M^{2}\over \tilde{r}^{2}}\left( - dt^{2}\,+\,d\tilde{r}^{2}\,
+\,\tilde{r}^{2} d\theta^{2}\,+\,
\tilde{r}^{2} \sin^{2} \theta d\phi^{2}\right),
\end{equation}
as could be  demonstrated [28] expanding 
the line element (55)
in power series about the event horizon
and subsequently making substitution 
\begin{equation}
r\,=\,M \left( 1\,+\,{M\over \tilde r}\right).
\end{equation}
The stress-energy tensor of the massive fields in the Bertotti-Robinson
may be easily obtained either by constructing the curvature terms for the line
element (55) and inserting them
into (25) or taking $|e|\,=\,M$ limit in the $\langle T^{\mu}_{\nu}\rangle^{(s)}_{ren}$
near the event horizon.
Simple calculations  give~\cite{notka}
\begin{equation}
\langle T^{\mu}_{\nu}\rangle^{(s)}_{ren}\,=\,{\mu^{(s)}\over 2880 \pi^{2} m^{2} M^{6}}
{\rm diag}[1, 1, -1, -1],
\end{equation}
where
\begin{equation}
\mu^{(s)}\,=\, 
\left\{\begin{array}{l}%\bigskip
{16\over 21} - 4 (\xi - {1\over 6})\\
%\bigskip
{37\over 14}\\
%\bigskip	  
{114\over 7}
\end{array}\right. .
\end{equation}

Assuming that the renormalized cosmological constant, $\Lambda_{ren},$
is zero in the analog of the  gravitational action (7) with the renormalized bare
lambda coefficients,
the Bertotti-Robinson geometry is a self-consistent solution of the 
semiclassical Einstein equations  with the source term given by the stress-energy tensor
of the massive field  in the large mass limit~\cite{Varun} if $\mu^{(s)}\,<\,0.$
It is because 
\begin{equation}
H^{\mu \nu}\,=\,{1\over g^{1/2}} {\delta \over \delta g_{\mu \nu}}
\int d^{4}x g^{1/2} R^{2},
\end{equation}
and
\begin{equation}
I^{\mu \nu}\,=\,{1\over g^{1/2}} {\delta \over \delta g_{\mu \nu}}
\int d^{4}x g^{1/2} R_{\mu \nu} R^{\mu \nu},
\end{equation}
vanish for the line element (55).
An interesting consequence of (57) is that a self-consistent solution is
possible for the massive scalar field provided $\xi\,>\,5/14,$  whereas
for massive spinor and vector fields appropriate solutions do not exist.
It should be noted however, that 
the stress-energy tensor of the massive
scalar field with the physically most plausible values 
of the coupling constant, namely $\xi = 0$ and $\xi =1/6,$
do not yield self-consistent solutions and therefore the scalar field
is not different than the spinor or vector field in this regard.

\section{Concluding remarks}

 In this paper we have constructed the renormalized stress-energy tensor
 of the massive scalar, spinor, and vector fields in the Reissner-Nordstr\"om
 spacetime. The method employed here is based on observation that the first
 order effective action  could be expressed  in terms of the traced coincidence limit
 of the coefficient $a_{3}.$ The general $ \langle T^{\mu}_{\nu}\rangle^{(s)}_{ren},$
 that has been obtained by functional differentiation of the effective action
 with respect to a metric tensor, consists of over one hundred terms,
 such as the terms cubic in curvature or involving fourth derivatives.
 Since even after simplifications, the final result is rather complicated
 the specific calculations are long but straightforward.
 
 Applying Eqs.(26-35) to the massive scalar field we rederived the results of Anderson,
 Hiscock, and Samuel.  Their calculations were based on the WKB
 approximation of the solutions of the scalar field equation 
 and  summation thus obtained  mode functions by means of the 
 Abel-Plana formula. On the other hand, the method employed here may be regarded as 
 geometrical and the identity of results is, although expected, impressive.
 To our knowledge spinor and vector fields have not been discussed earlier.
 
 The results (39-51) have also been used to construct and analyse stress-energy
 tensor in the two interesting limiting cases that could be obtained
 from the Reissner-Nordstr\"om solution: the extremal Reissner-Nordstr\"om 
 and  Bertotti-Robinson geometries. 
 Because of the form of the stress-energy tensor and the fact that
 the variational derivatives of  the functionals constructed from
 $R_{\mu \nu} R^{\mu \nu}$	and $R^{2}$ vanish in the Bertotti-Robinson spacetime,
this geometry may be a self-consistent solution of the
 semiclassical Einstein field equations.
We found that the self-consistent solution is
possible for the massive scalar field provided $\xi\,>\,5/14,$ 
whereas for massive spinor and vector fields such solutions do not exist.

Finally, we remark that it would be interesting  to construct the next order of the renormalized
effective action (4). As the functional 	 $W_{ren}$ at that order involves
coincidence limit of the $a_{4}$ coefficient, which is, in turn, given
by a very complicated formula, one expect that such a calculation would be
a real challenge. 			  
Another important direction of investigation is
generalization of the  obtained results to the elliptic operators (1) with
other physically interesting matrix potentials $Q$	and curvatures $ {{\cal R}^{A}}_{B},$
and to analyse the back reaction of the quantized massive fields  on the metric.
We hope that the results obtained in this paper will be of use in further
calculations.

It should be emphasized however, that 
being local in its nature, the Schwinger-DeWitt expansion  does not describe
particle creation which is a
nonperturbative and nonlocal phenomenon. Moreover, in the massless limit
the method breaks down.
To address successfully this group of problems
new methods are necessary as, for example, the covariant perturbation theory \cite{vilk2}.

 \begin{table}
 \caption{The coefficients $\al_{i}^{(s)}$ for the massive scalar, spinor, and vector
 field}
\begin{tabular}{cccc}
& s = 0 & s = 1/2 & s = 1 \\ 
$\al^{(s)}_{1} $ & $ {1\over2}\xi^{2}\,-\,{1\over 5} \xi $\,+\,${1\over 56}$ & $- {3\over 140}$ &
$- {27\over 280}$\\	  
 $\al^{(s)}_{2}$ & ${1\over 140}$ & ${1\over 14}$ & ${9 \over 28}$\\ 
 $\al^{(s)}_{3}$ &$ \left( {1\over 6} - \xi\right)^{3}$& ${1\over 432}$ &$ - {5\over 72}$\\ 
 $\al^{(s)}_{4}$ & $- {1\over 30}\left( {1\over 6} - \xi\right) $& $- {1\over 90}$ & ${31\over 60}$\\
 
 $\al^{(s)}_{5}$ & ${1\over 30}\left( {1\over 6} - \xi\right)$ &$ - {7\over 720}$ &$ - {1\over 10}$\\
 
 $\al^{(s)}_{6}$ & $ - {8\over 945}  $& $- {25 \over 376}$ & $- {52\over 63}$\\
 
 $\al^{(s)}_{7}$ & ${2 \over 315}$ & $ {47\over 630}$  & $- {19\over 105} $\\	
 $\al^{(s)}_{8}$ & ${1\over 1260}$ & ${19\over 630} $ & ${61\over  140} $\\	  
 $\al^{(s)}_{9}$ & ${17\over 7560}$& ${29\over 3780}$ & $- {67\over 2520}$\\	  
 $\al^{(s)}_{10}$ & $- {1\over 270}$ & $ - {1\over 54} $  & $ {1\over 18}$\\	  
 \end{tabular}
 \label{table1}
 \end{table}

%%%%%%%%%%%%%%% FIGURES %%%%%%%%%%%%%%%%%%%%%%%%%
%%%%%%%%%%%%%%%%%%%%%%%%%%%%%%%%%%%%%%%%%%%%%%%%%

\begin{figure}[htb]
\vbox{\hfil\epsfbox{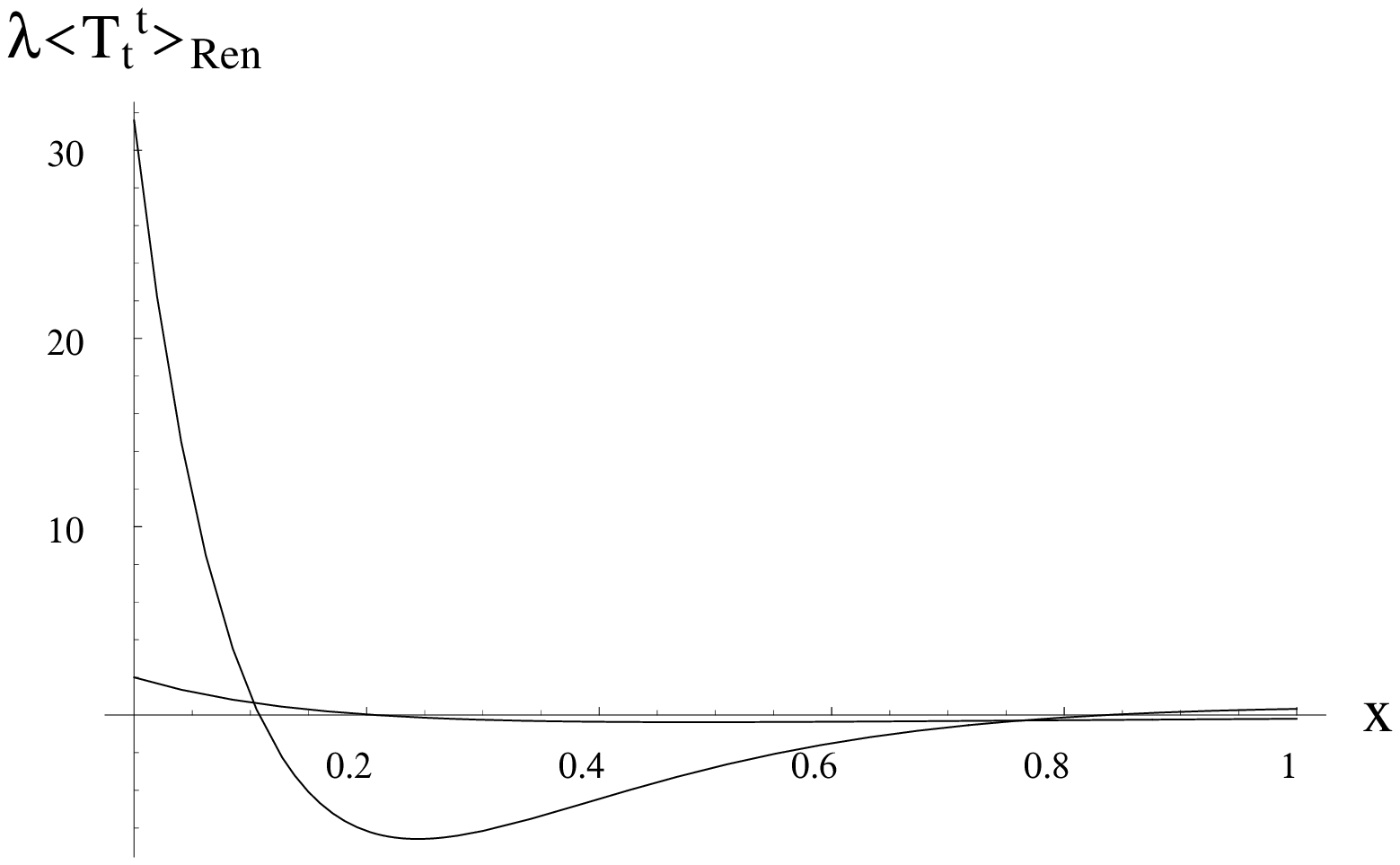}\hfil}

\caption[T]
 {\label{fig1}This graph shows the radial dependence of the 
 rescaled component
 $\langle T^{t}_{t}\rangle_{ren}^{(1/2)}$	 $(\lambda\,=\,180 (8 M)^{4} \pi^{2})$
 of the renormalized stress-energy tensor of the massive spinor field with 
 $m\,=\,2/M	.$ From top to bottom the cureves are for $q = 0.95$ and $q = 0.$

} 
\end{figure}
\clearpage
%%%%%%%%%%%%%%%%%%%%%%

\begin{figure}[htb]
\vbox{\hfil\epsfbox{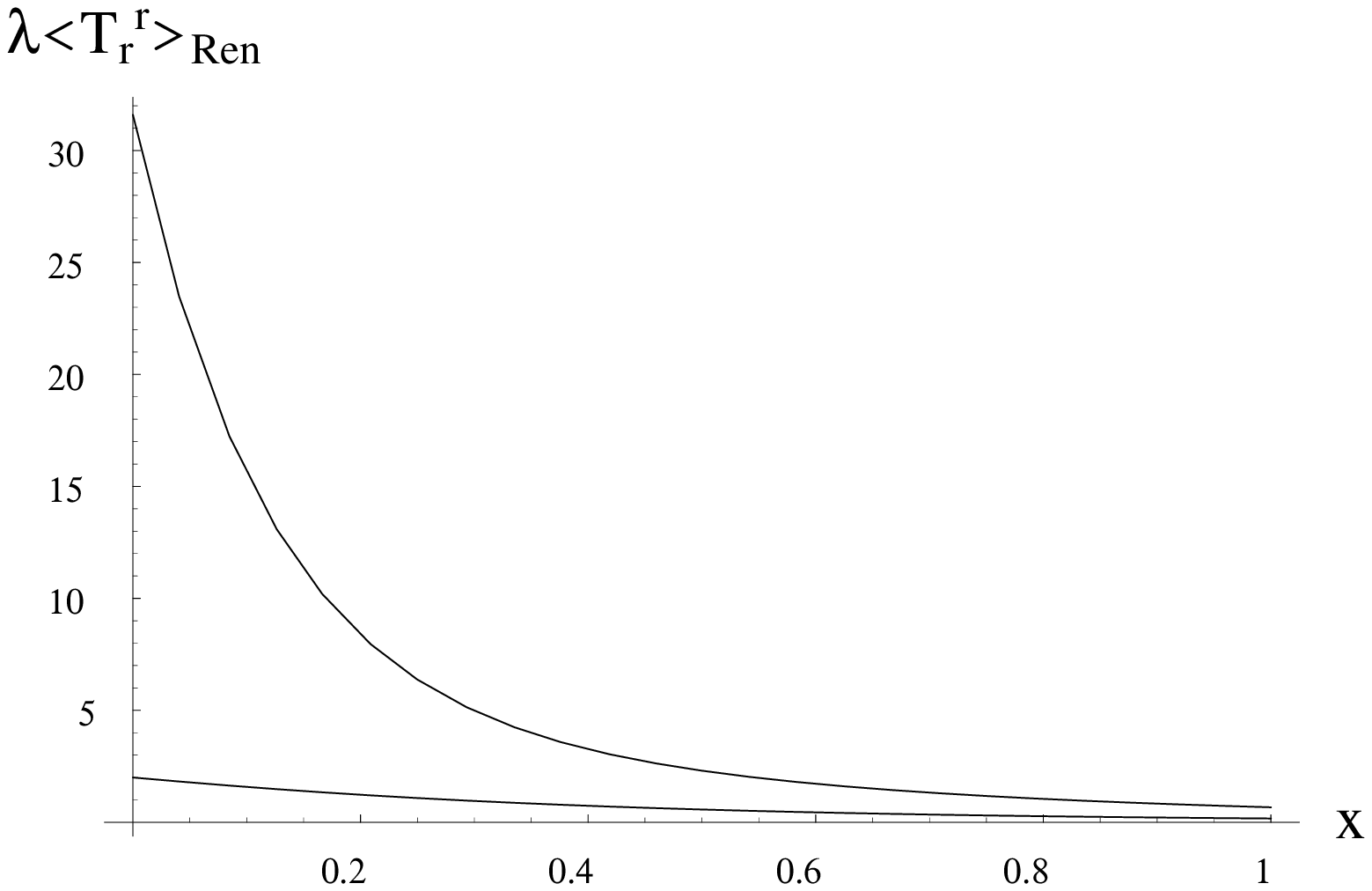}\hfil}

\caption[T]
 {\label{fig2}This graph shows the radial dependence of the 
 rescaled component
 $\langle T^{r}_{r}\rangle_{ren}^{(1/2)}$	 $(\lambda\,=\,180 (8 M)^{4} \pi^{2})$
 of the renormalized stress-energy tensor of the massive spinor field with 
 $m\,=\,2/M	.$ From top to bottom the cureves are for $q = 0.95$ and $q = 0.$ } 
\end{figure}
 \clearpage
 %%%%%%%%%%%%%%%%%%%%%

\begin{figure}[htb]
\vbox{\hfil\epsfbox{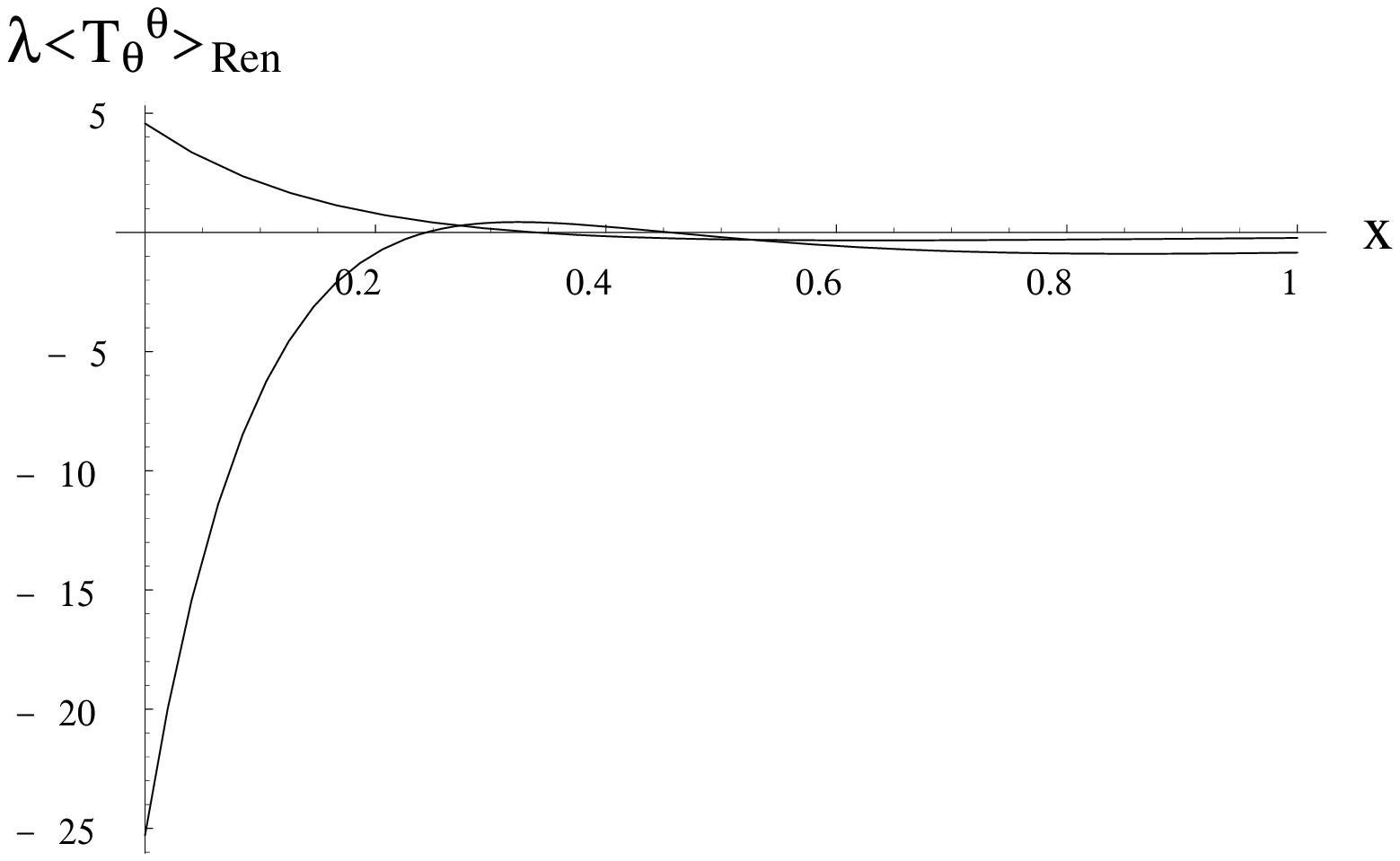}\hfil}

\caption[The fractional trace anomaly]
 {\label{fig3}This graph shows the radial dependence of the 
 rescaled component
 $\langle T^{\theta}_{\theta}\rangle_{ren}^{(1/2)}$	 $(\lambda\,=\,180 (8 M)^{4} \pi^{2})$
 of the renormalized stress-energy tensor of the massive spinor field with 
 $m\,=\,2/M	.$ From top to bottom the cureves are for $q = 0$ and $q = 0.95$ } 
\end{figure}
\clearpage
%%%%%%%%%%%%%%%%%%%%%

\begin{figure}[htb]
\vbox{\hfil\epsfbox{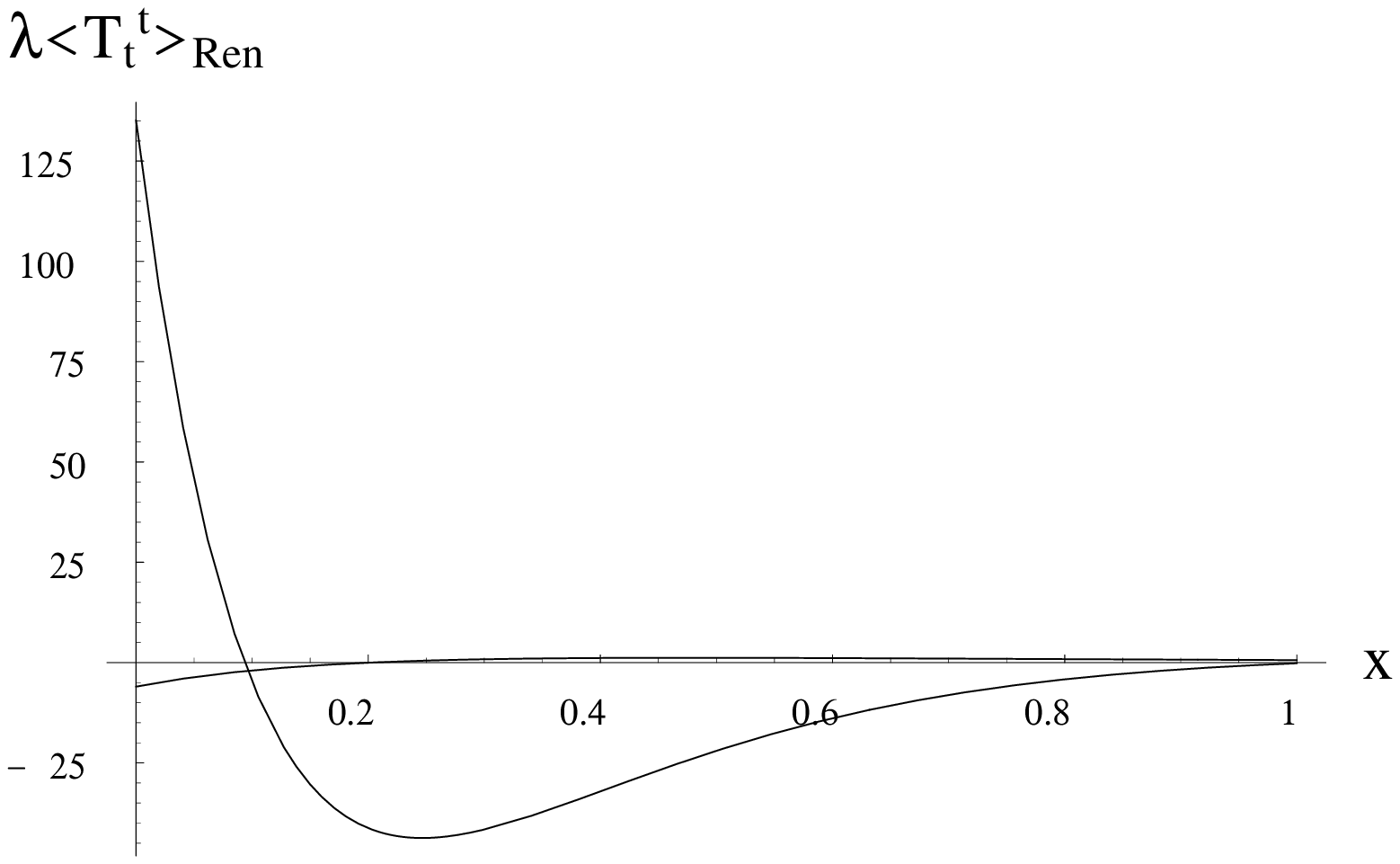}\hfil}

\caption[The fractional trace anomaly]
 {\label{fig4}This graph shows the radial dependence of the 
 rescaled component
 $\langle T^{t}_{t}\rangle_{ren}^{(1)}$	 $(\lambda\,=\,90 (8 M)^{4} \pi^{2})$
 of the renormalized stress-energy tensor of the massive vector field with 
 $m\,=\,2/M	.$ From top to bottom the cureves are for $q = 0.95$ and $q = 0.$ } 
\end{figure}
\clearpage
%%%%%%%%%%%%%%%%%%%%%%

\begin{figure}[htb]
\vbox{\hfil\epsfbox{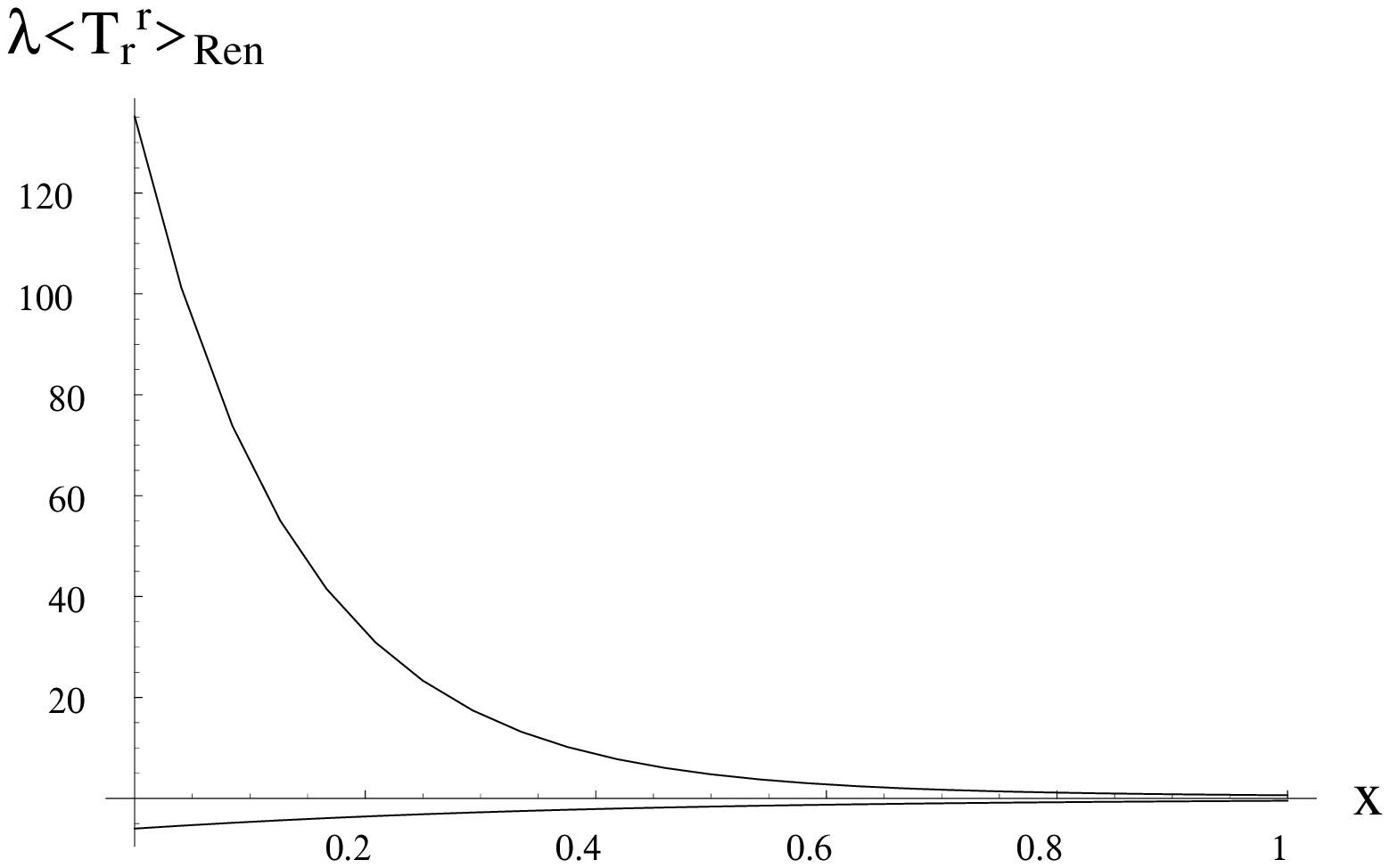}\hfil}

\caption[The fractional trace anomaly]
 {\label{fig5}This graph shows  the radial dependence of the 
 rescaled component
 $\langle T^{r}_{r}\rangle_{ren}^{(1)}$	 $(\lambda\,=\,90 (8 M)^{4} \pi^{2})$
 of the renormalized stress-energy tensor of the massive vector field with 
 $m\,=\,2/M	.$ From top to bottom the cureves are for $q = 0.95$ and $q = 0.$} 
\end{figure}
\clearpage
%%%%%%%%%%%%%%%%%%%%%%

\begin{figure}[htb]
\vbox{\hfil\epsfbox{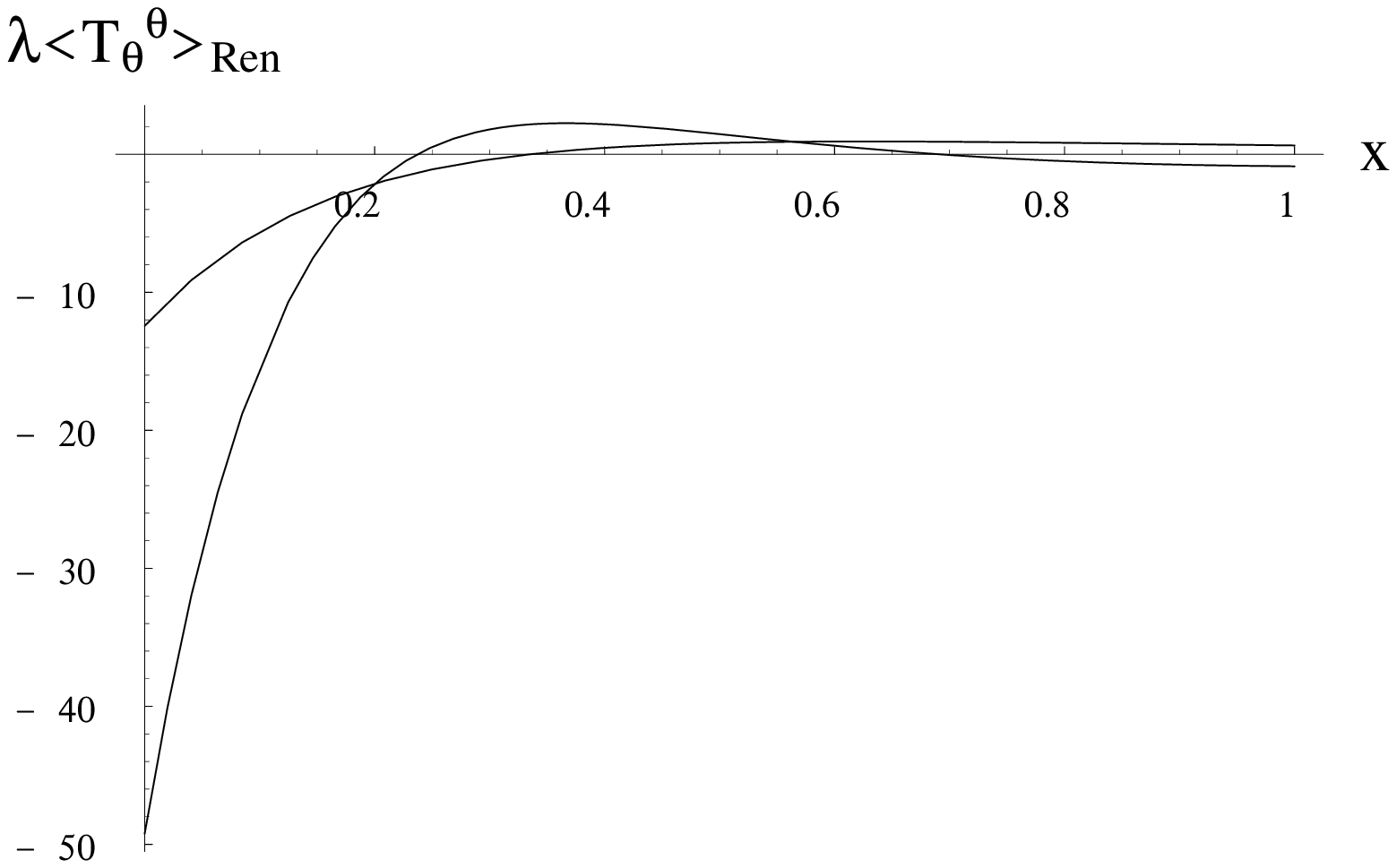}\hfil}

\caption[The fractional trace anomaly]
 {\label{fig6}This graph shows  the radial dependence of the 
 rescaled component
 $\langle T^{\theta}_{\theta}\rangle_{ren}^{(1)}$	 $(\lambda\,=\,90 (8 M)^{4} \pi^{2})$
 of the renormalized stress-energy tensor of the massive vector field with 
 $m\,=\,2/M	.$ From top to bottom the cureves are for $q = 0$ and $q = 0.95$} 
\end{figure}
\clearpage


\begin{references}
\bibitem{notka} Our $\mu^{(0)}$ differs from the value evaluated 
in Ref.~\cite{Varun} which is insenstive to the coupling constant $\xi.$
\end{references}
\end{document}